\documentclass[12pt]{iopart}
\usepackage[dvips,final]{graphicx}
\usepackage[defs]{ams}
\usepackage{url}

\usepackage{bm}

\usepackage{color}
\usepackage{ulem}

\begin{document}
\title{Hydrogen mediated ferromagnetism in ZnO single crystals}

\author{M. Khalid,
P. Esquinazi}
\address{ Division of Superconductivity and Magnetism,
University of Leipzig, D-04103 Leipzig, Germany}
\author{D. Spemann}
\address{ Division of Nuclear Solid State Physics, University of
Leipzig, D-04103 Leipzig, Germany}
\author{W. Anwand, G. Brauer}
\address{ Institut f{\"u}r Strahlenphysik, Forschungszentrum
Dresden-Rossendorf, PO Box 510119, D-01314 Dresden, Germany}
\date{\today}

\begin{abstract}
We investigated the magnetic properties of hydrogen plasma treated
ZnO single crystals by SQUID magnetometry.  In agreement with the
expected hydrogen penetration depth we found ferromagnetic
behavior located at the first 20~nm of the H-treated surface of
ZnO
 with magnetization at saturation up to
6~emu/g at 300~K and Curie temperature T$_c$ $\gtrsim $ 400~K. In
the ferromagnetic samples a hydrogen concentration of a few atomic
percent in the first 20~nm surface layer was determined by nuclear
reaction analysis. The saturation magnetization of H-treated ZnO
increases with the concentration of hydrogen.
\end{abstract}
\pacs{75.70.-i, 75.20.Ck, 75.30.Hx, 75.50.Pp}

\submitto{\NJP}

\maketitle

After a large number of studies and different kinds of efforts,
experimental and theoretical work of the last years indicate that
defect-induced magnetism remains the key to trigger ferromagnetism
in ZnO (as well as in other non-magnetic oxides) with Curie
temperature above 300~K. Not the doping with magnetic elements
appears to be a successful and reproducible method to trigger
magnetic order in this oxide but the introduction of a certain
defect density of the order of a few percent, like O- \cite{Ban07}
or Zn-vacancies \cite{kha09} with or without doping of
non-magnetic ions like C \cite{Zhou08}, N \cite{Wu10},
Li\cite{Cha09} or Cu \cite{Xu08,Yi10}. In general, however, the
achieved magnetization values are still too low, indicating that
the magnetic order is very probably not homogeneously distributed
in the whole sample, a necessary condition for application of this
phenomenon in ZnO-based devices.

What about the influence of hydrogen in the magnetism of ZnO? It
is known that the presence of hydrogen  is unavoidable in all
systems and in general it remains rather difficult to measure it
with high enough accuracy. Hydrogen related magnetic order was
recently found in graphite surfaces by x-ray magnetic circular
dichroism \cite{ohl10} indicating that this element can play a
role in the magnetism of nominally non magnetic materials. The
role of hydrogen in ZnO can be diverse. It can act either as donor
(H$^+$) or acceptor (H$^-$) and can even modify the host
structure. In bulk ZnO hydrogen acts as a shallow donor and is a
source of the unintentional n-type conductivity \cite{Wal00}. Room
temperature ferromagnetism due to atomic hydrogen adsorption on
different terminated surfaces of ZnO \cite{Wang08,Wol07,San10} or
in the bulk of Co-doped ZnO \cite{liu09} has been studied
theoretically. In one of these studies it was shown that atomic
hydrogen adsorbed on the Zn place on the ZnO(0001) surface forms a
strong H-Zn bonds leading to a metallic surface with a net
magnetic moment \cite{San10}. All these theoretical studies
indicate that it is important to check whether hydrogen
implantation can trigger ferromagnetism in ZnO. In this work we
investigate the magnetic properties of ZnO single crystals treated
by remote hydrogen plasma and demonstrate that a remarkable
ferromagnetic signal with a large magnetization is located in a
few nm of the H-treated surface of the ZnO single crystals. Our
finding opens up a simple and reproducible possibility to trigger
magnetic order in broad or localized regions of ZnO bulk, thin
films or microstructures without the need of introducing other
elements or vacancies.

Hydrothermally grown ZnO single crystals were used for H-plasma
treatment. Two of them were with one-side polished (O-terminated)
of dimensions $\sim$ $(10\times10\times0.5)$ mm$^3$ (MaTeck GmbH,
J\"ulich). One of them was treated with H-plasma while the other
one was kept as reference. Both single crystals  were cut before
treatment to directly mount them in a straw for the magnetic
measurements done with  a Superconducting Quantum Interferometer
Device (SQUID). Four other crystals were two sides polished with
similar termination but of dimensions $\sim
(6\times6\times0.5)$~mm$^3$ (CrysTec GmbH, Berlin). Hydrogen
doping in ZnO can be achieved by, e.g. adsorption of hydrogen on
the ZnO surface, H-implantation or remote hydrogen plasma doping
\cite{Tor09,Wan09,Str04}. We used the last method to implant
hydrogen into ZnO.

\begin{figure}
\begin{center}
\includegraphics[width=0.9\linewidth]{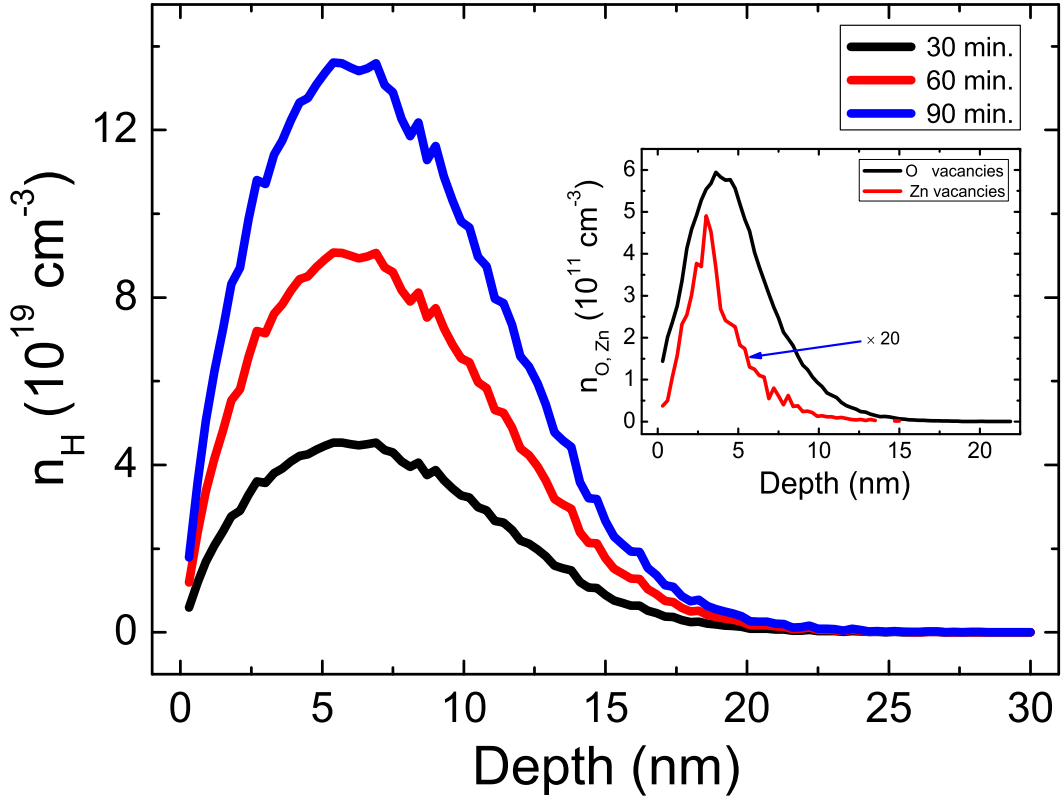}
\caption{\label{etc} Hydrogen concentration vs penetration depth
of H$^+$-ions in ZnO single crystals at different total implanted
time, estimated by SRIM. The simulation results show that most of
the H$^+$-ions are implanted
 within the first 20~nm from the surface. The inset shows the oxygen and zinc
vacancies concentration produced in ZnO single crystals during 90
minutes H-implantation at the used energy conditions. }
\end{center}
\end{figure}

There are several parameters that may influence the strength of
the magnetic order triggered through H-plasma treatment. Namely,
the sample temperature, the $H^+$-ion energy and current and the
total implanted charge. In this work we demonstrate how the
substrate temperature and the total implanted charge  influence
the magnetic ordering in ZnO crystals. The substrate temperature
was varied from room temperature to $400\,^{\circ}{\rm C}$ whereas
the total implanted charge was controlled by varying the treatment
time ranging from 30 to 90 min. The ZnO surface was placed
$\sim$100~nm downstream from the plasma with a bias voltage of
$\sim$-330 V (parallel-plate system of voltage difference of
$10^3$~V). The bias current was fixed at $\sim$ 50 $\mu$A (sample
plus sample holder) while the current into the sample only was nA.
The exposition to remote hydrogen dc plasma ranges from 30 to 90
min. During the loading the gas pressure was $\sim$1~mbar.
\begin{figure}
\begin{center}
\includegraphics[width=0.9\linewidth]{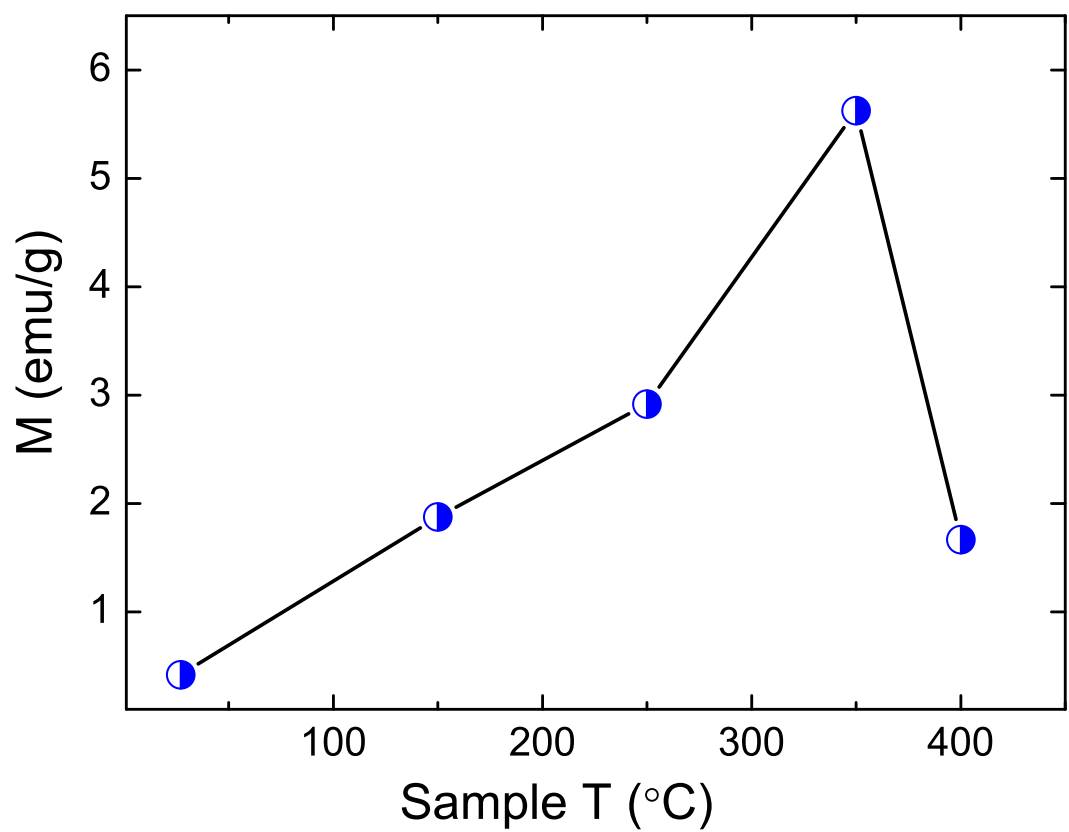}
\caption{\label{hys} Room temperature saturation magnetization of
the ferromagnetic signal of H-ZnO single crystals treated at
different substrate temperatures, all with similar nominally
implanted charge.}
\end{center}
\end{figure}

\begin{figure}
\begin{center}
\includegraphics[width=0.9\linewidth]{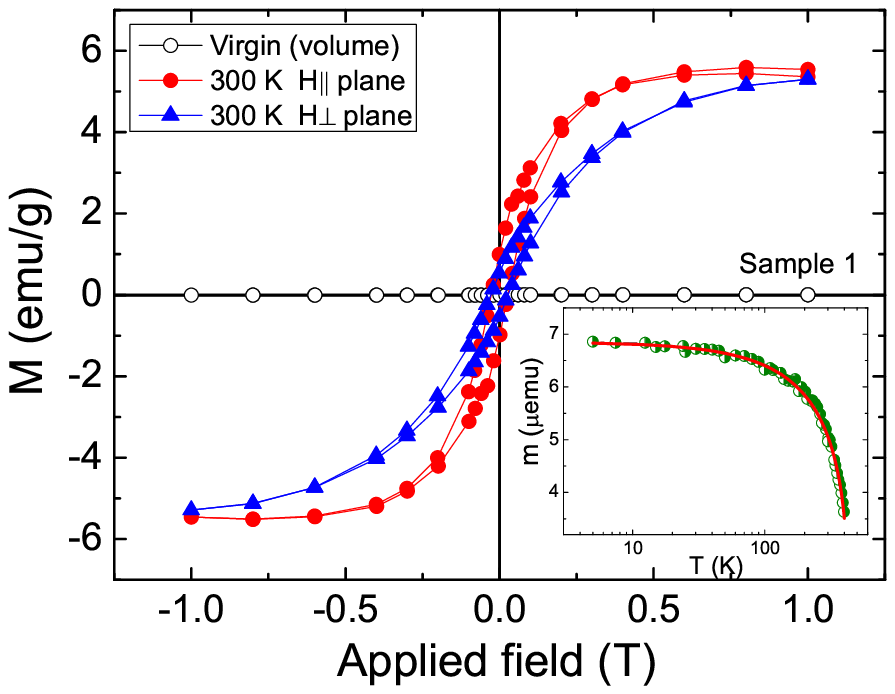}
\caption{\label{etc}Hysteresis loops of a H-ZnO single crystal
measured at 300~K applying a magnetic field parallel $(\bullet)$
and perpendicular $(\blacktriangle)$ to the main plane of the
sample. The open circles $(\circ)$ represent the ferromagnetic
magnetization of the untreated ZnO crystal calculated by taking
into account the whole volume of the sample. The diamagnetic
linear background was subtracted from the measured signal. The
inset shows the remanent moment vs. temperature.  The (red) solid
line is given by  $m(T)= 6.8 [\mu$emu$] (1-T/T_c)^{1/3})$ with a
Curie temperature $T_c = 450~$K.}
\end{center}
\end{figure}

The hydrogen content before and after H-treatment was determined
by standard Nuclear Reaction Analysis (NRA) \cite{Lan95} using
6.64~MeV $^1$$^5$N ions with a depth resolution of $\sim$5~nm and
a H detection limit of $\sim$200~ppm. The H-concentration vs.
depth was obtained using SRIM (The Stopping and Range of Ions in
Matter) simulation \cite{Zie85}, see Fig.~1. The average hydrogen
atomic concentration of bound hydrogen measured by NRA within the
top 100~nm surface region of the ZnO crystals  before and after
60~min treatment was (0.14$\pm$0.03) and (0.64$\pm$0.07) at.$\%$,
respectively. Comparable results were reported recently
\cite{anw10}. The average H-concentration reached in the 20~nm
surface region after 60 min implantation at the used current
conditions was $\sim 2.5 \pm 0.5~$at.\%. Particle induced x-ray
emission (PIXE) measurements were used to analyze the magnetic
impurities of the samples before and after H-treatment. There was
no significant difference in the magnetic impurity concentration
before and after H-plasma treatment.

Several ZnO single crystals were treated with H-plasma. All of
them showed an increase in the ferromagnetic moment  after
H-plasma treatment relative to their virgin values. The increase
of the ferromagnetic moment depends on the temperature of the
sample during H-treatment at nominally similar total implanted
charge. Figure~2 shows the saturation values of the ferromagnetic
magnetization signal (after subtraction of the diamagnetic linear
background) of several samples treated in H-plasma for 90~min at
different temperatures. We found that the ferromagnetic saturation
magnetization increases by increasing sample temperature and
reaches a maximum at $350\,^{\circ}{\rm C}$. Therefore we
concentrate on the study of samples implanted at
$250\,^{\circ}{\rm C}$ and $350\,^{\circ}{\rm C}$. We note that
the observed dependence on the sample temperature might be very
useful to control H diffusion as well as lattice defects produced
during plasma treatment as well as after treatment.

In what follows we discuss results of samples treated at
$350\,^{\circ}{\rm C}$ and $250\,^{\circ}{\rm C}$ for 1~h
implantation and followed by results of samples implanted at
 $350\,^{\circ}{\rm C}$ but with different
total charge (or implantation time). The results shown in this
paper were not affected by aging after leaving the samples one
year at 300~K.

\begin{figure}
\begin{center}
\includegraphics[width=0.9\linewidth]{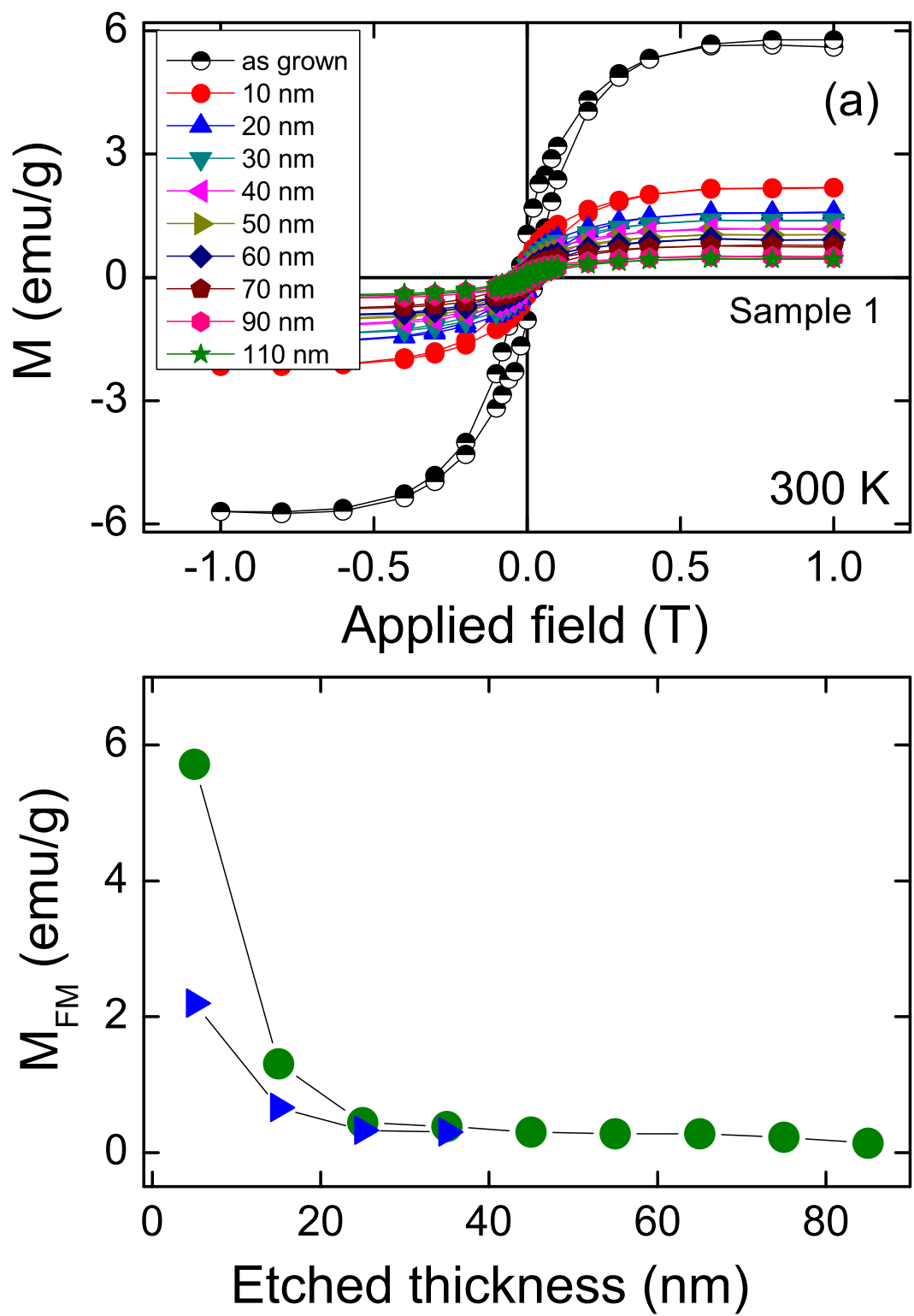}
\caption{\label{etc} (a) Magnetic moment as a function of applied
magnetic field of a H-ZnO single crystal at different etching
stages, after subtraction of the diamagnetic contribution at
300~K. (b) Ferromagnetic magnetization values at saturation
obtained taking into account the change in magnetic moment (see
(a)) after etching a specific thickness of the surface region for
H-treated samples  at substrate temperatures of $350\,^{\circ}{\rm
C}$ ($\bullet$) and
 $250\,^{\circ}{\rm C}$ $(\blacktriangle)$.}
\end{center}
\end{figure}

Figure 3 shows the magnetic moment at 300~K of a H-ZnO single
crystal treated at $350\,^{\circ}{\rm C}$ and at applied magnetic
fields parallel and perpendicular to the main sample area. Clear
ferromagnetic hysteresis loops are observed at 300~K. The
ferromagnetic behavior of the hysteresis shows a clear
magnetocrystalline anisotropy with anisotropy constant
$\textit{K}$$_1$ $\sim$ 2 $\times$ 10$^5$ J/m$^3$. This anisotropy
also excludes magnetic impurities as origin for the observed
ferromagnetism. The remanent magnetic moment vs. temperature shown
in the inset of Fig.~3 was measured at zero field after applying a
field parallel to the sample main surface and cooled down in field
to 5~K. The temperature dependence of the remanence follows
$m(T)=m_0(1-T/T_c)^{\delta}$ with $\delta = 0.33 \pm 0.05$, a
static scaling law with an exponent similar to other ferromagnets
like e.g. CrBr$_3$\cite{ho70}. This fit indicates a Curie
temperature of $T_c = 450~\pm 25~$K.

In order to investigate how much surface thickness of the ZnO
single crystals is contributing to the observed ferromagnetism  we
etched up to $\sim 100$~nm thick layer and studied the change in
ferromagnetic moment.  For this purpose we used  a solution of
0.3~ml HCl in 400~ml water\cite{Mak02,Lin77}. The single crystals
were etched from both sides and then the measured etched mass
divided by two in order to exclude the mass of H-untreated side.
After $\simeq 40~$s etching time a $\simeq 4~\mu$g mass of the ZnO
crystal was removed, which means $\simeq$10~nm thick layer from
the H-treated side. After etching 10~nm thick layer the hysteresis
loop was measured at 300~K. These results are shown in Fig. 4(a).
Knowing the etched mass and the corresponding change in
ferromagnetic moment we can calculate the real magnetization of
the H-treated layer. The magnetization as a function of etched
thickness for two samples treated at different temperatures is
shown in Fig. 4(b). From these results it becomes clear that the
major ferromagnetic contribution vanishes after etching the first
$\simeq$20~nm layer, a layer thickness that agrees with the
calculated H-concentration using SRIM and shown in Fig.1. At the
used $H^+$-energies the estimated concentration of O- and
Zn-vacancies is 8 and 9 orders of magnitude smaller than the
H-concentration in the first 20~nm from the surface, see inset in
Fig.~1. This huge difference clearly indicates that in the treated
samples hydrogen with a concentration of several percent at the
surface  and not Zn- or O-vacancies or interstitials play a major
role in the observed magnetic order.

\begin{figure}
\begin{center}
\includegraphics[width=0.9\linewidth]{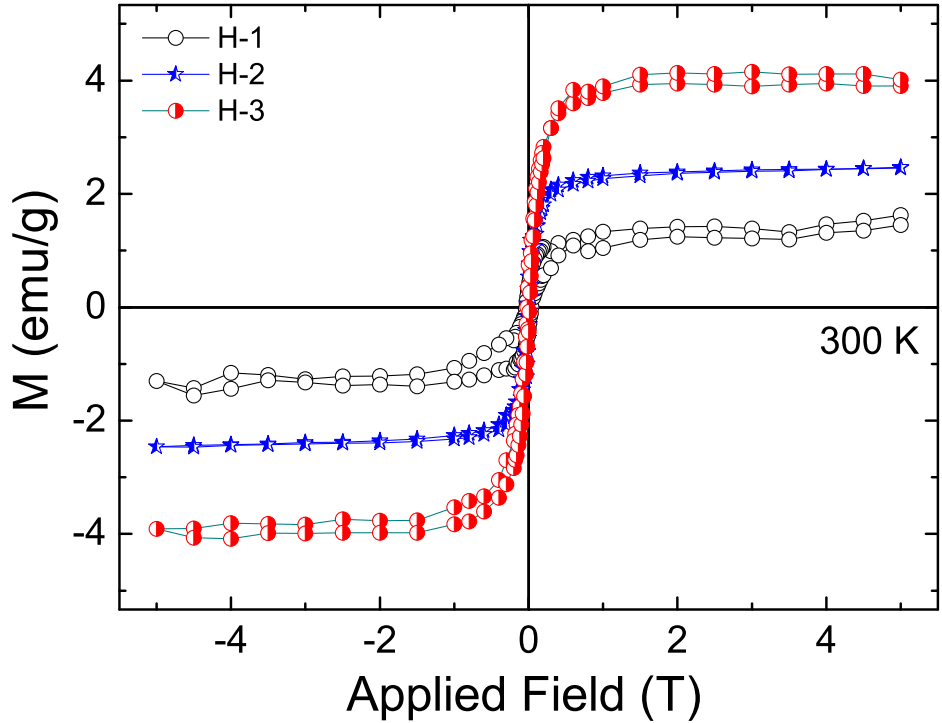}
\caption{\label{mr}  Magnetization of H-ZnO single crystal vs
applied field measured at different total implantations. The
saturation magnetization increases by increasing the implanted
charge.}
\end{center}
\end{figure}

We also studied the influence on the ferromagnetic signal of the
amount of hydrogen implanted into the sample. The samples were
treated with H-plasma at a temperature of $350\,^{\circ}{\rm C}$
at 3 different total implanted times, namely 30, 60 and 90 min.
The ferromagnetic signals  of these samples are shown in Fig.~5.
We observe that the magnetization of H-ZnO crystals increases with
the total treatment time.

With the measured ferromagnetic magnetization within the first
20~nm surface region we estimate a magnetic moment of the order of
$0.2~\mu_B$ per ZnO unit cell. If we assume that in average about
1~H atom per unit cell exists in this 20~nm region then this
magnetic moment triggered by each H atom is comparable to that
obtained in Ref.~\cite{San10}.\\

In conclusion, we investigated the magnetic properties of remote
H-plasma treated ZnO single crystals. The NRA results confirmed
the enhanced concentration of hydrogen in ZnO single crystals
after treatment. Characteristic ferromagnetic hysteresis loops as
well as a magnetic anisotropy were observed in H-ZnO samples at
room temperature. Systematically measurements of the magnetic
moment of the treated samples after wet chemical etching proved
that only the first $\lesssim$ 20~nm thick surface layer of
H-treated ZnO contributes to the total ferromagnetic
magnetization, in agreement with the expected H-penetration depth
estimated by SRIM. We attribute the observed ferromagnetism in
H-ZnO single crystals to the influence of hydrogen. Because
hydrogen implantation also reduces dramatically the resistivity of
the ZnO structure, this ferromagnetic oxide should be more easily
applied in spintronic devices. Magnetotransport measurements on
similar H-treated samples  are currently being performed and show
a negative magnetoresistance that increases (absolutely speaking)
the larger the ferromagnetic magnetization (proportional to the
H-concentration). Transport measurements as a function of the
angle between current and applied magnetic field show a clear
anisotropic magnetoresistance (AMR), which amplitude decreases as
a function of temperature. At 250~K the amplitude of the AMR is
$\simeq 0.5\%$ of the measured resistance at a field of 5~T. The
existence of an AMR indicates a finite $L \cdot S$ coupling
contribution to the scattering as well as the existence of
a spin asymmetry in the electronic band. These results will be published elsewhere.\\

This work was supported by the DFG within the Collaborative
Research Center (SFB 762) ``Functionality of Oxide Interfaces''.\\

\bibliographystyle{unsrt}

\end{document}